\documentclass[10pt,journal]{IEEEtran}

\usepackage{cite}
\usepackage{mathtools}
\usepackage{amsmath,amssymb,amsfonts}
\usepackage{algorithmic}
\usepackage{tikz}
\usepackage{pgfplots}
\usepackage{pgfplotstable}
\usepackage{booktabs}
\usetikzlibrary{matrix, positioning}
\usepgfplotslibrary{statistics}
\pgfplotsset{compat=1.16}
\usepackage{textcomp}
\usepackage{hyperref}
\usepackage{cleveref}
\usepackage[usestackEOL]{stackengine}

\newcommand{\R}{\ensuremath{\mathbb{R}}}

\newcommand{\vx}{\ensuremath{\mathbf{x}}}
\newcommand{\vz}{\ensuremath{\mathbf{z}}}
\newcommand{\sdfvec}{\ensuremath{\mathbf{s}}}

\newcommand{\latentspace}{\ensuremath{\mathbb{R}^d}}
\newcommand{\latentcode}{\ensuremath{\vz}}
\newcommand{\nnweights}{\ensuremath{\theta}}

\newcommand{\shapeopmulti}{\ensuremath{\mathbf{S}_{\nnweights}}}

\newcommand{\imgpartial}[3]{%
    \includegraphics[width=1.4cm,trim={200 20 220 20},clip]{figures/partial_data/mesh_#1_points_#2_#3.png}
}

\begin{document}

\title{Uncertainty Quantification for Cardiac Shape Reconstruction with Deep Signed Distance Functions via MCMC methods}
\author{Jan Verhülsdonk, Thomas Grandits, Francisco Sahli Costabal, Thomas Beiert, Simone Pezzuto, Alexander Effland
\thanks{This work was partially funded by the Deutsche Forschungsgemeinschaft (DFG, German Research Foundation) -- CRC 1720 -- 539309657, Germany's Excellence Strategy 
EXC 2047 -- 390685813, Hausdorff Center for Mathematics and EXC 2151 -- 390873048. SP and TG are supported by the SNSF project ``CardioTwin'' (no.~214817). SP acknowledges the support of the CSCS-Swiss National Supercomputing Centre project no.~lp100 and the PRIN-PNRR project no.~P2022N5ZNP. SP is member of INdAM-GNCS.}
\thanks{J.~Verh\"ulsdonk and A.~Effland are with the In\-sti\-tu\-te for App\-lied Ma\-the\-ma\-tics, University of Bonn, Germany (e-mail: \{verhuelsdonk,pinetz,effland\}@iam.uni-bonn.de).}
\thanks{T.~Grandits is with the Department of Mathematics and Scientific Computing, University of Graz, Austria and with NAWI Graz (e-mail: thomas.grandits@uni-graz.at).}
\thanks{F.~Sahli Costabal is with the Institute for Biological and Medical Engineering, Pontificia Universidad Católica de Chile, Chile (e-mail: fsahli1@uc.cl).}
\thanks{T.~Beiert is with the Heart Centre Bonn, Department of Medicine II, University Hospital Bonn, Germany (e-mail: thomas.beiert@ukbonn.de).}
\thanks{S.~Pezzuto is with the Laboratory of Mathematics for Biology and Medicine, Department of Mathematics, University of Trento, Italy,
and with the Euler Institute, Universit\`a della Svizzera italiana, Switzerland (e-mail: simone.pezzuto@unitn.it).}}

\maketitle

\begin{abstract}
Atlas-based approaches allow high-quality, patient-specific shape reconstructions of cardiac anatomy from sparse and/or noisy data such as point clouds. However, these methods are mainly prior-driven, so the impact of uncertainty can be large, limiting their clinical reliability. We propose a probabilistic framework for uncertainty-aware cardiac shape reconstruction that combines Deep Signed Distance Functions (DeepSDFs) with Markov Chain Monte Carlo (MCMC) sampling. Cardiac geometries are modeled implicitly as zero-level sets of a neural network conditioned on learned latent codes, enabling multi-surface reconstruction of the left and right ventricles. By interpreting the reconstruction loss as a log-likelihood, we perform Bayesian inference in the latent space to obtain both maximum a posteriori (MAP) and posterior-sampled reconstructions. Experiments on a public cardiac dataset show that our approach produces accurate reconstructions and well-calibrated uncertainty estimates.
\end{abstract}


\section{Introduction}
\label{sec:introduction}
Accurate reconstruction of cardiac anatomy is fundamental for understanding cardiovascular disease, guiding patient-specific therapy planning, and supporting risk stratification.
Such reconstructions are typically obtained from cardiac imaging modalities such as MRI and CT. To generate detailed cardiac geometries from these data, atlas-based approaches are often employed. These methods can bridge the gap between low-quality or sparse imaging data and detailed anatomical reconstructions suitable for numerical simulations.

To address challenges in geometric representation, including dependence on imaging modality and resolution as well as the need for efficient geometric regularization, Deep Signed Distance Functions (DeepSDFs)~\cite{park_deepsdf_2019,liu_learning_2022} represent shapes implicitly as the zero-level sets of neural networks, enabling smooth, continuous, and high-fidelity modeling of cardiac anatomy~\cite{khan, sander, verhulsdonk2024shape, kong}. Once a shape prior (atlas) has been trained, reconstruction amounts to solving a low-dimensional optimization problem to find the latent code that yields the best-fitting anatomy. Given the sparsity of the data in certain scenarios, e.g., electroanatomical mapping, the question of identifiability naturally arises. In this work, we address this challenge by recasting the inference problem in a Bayesian setting. We model the prior distribution in latent space as a Gaussian distribution inferred from the latent representations of cardiac anatomies in the training set, while the likelihood is determined by the data fidelity term.

For probabilistic inference in this setting, Markov chain Monte Carlo (MCMC) methods provide a principled framework for approximating posterior distributions in high-dimensional latent spaces. We employ Hamiltonian Monte Carlo (HMC) methods~\cite{neal2011mcmc,hoffman2014no} to sample latent shape codes conditioned on sparse or noisy point-cloud observations, yielding both the maximum a posteriori reconstruction and a distribution of plausible shapes that captures reconstruction uncertainty.

In this work, we focus on biventricular reconstructions that jointly model the left and right ventricular endocardial and epicardial surfaces. While prior DeepSDF-based approaches for anatomical reconstruction typically produce deterministic point estimates, they do not quantify uncertainty arising from sparse observations, noise, or shape ambiguity. We address this limitation by introducing a Bayesian inference framework for cardiac shape reconstruction.

Our main contributions are as follows:
\begin{itemize}
\item We develop, to the best of our knowledge, the first uncertainty quantification framework for cardiac shape reconstruction based on deep signed distance functions. To this end, we formulate latent shape inference as a posterior sampling problem and propose MCMC-based algorithms for uncertainty-aware reconstruction from sparse or noisy data.
\item We demonstrate that the proposed probabilistic formulation yields both accurate reconstructions and principled uncertainty estimates, enabling the assessment of prediction reliability and variability beyond conventional point-estimate methods.
\end{itemize}

\section{Related Work}
In the medical imaging domain, implicit neural representations such as DeepSDF have been increasingly applied to anatomical modeling. Khan et al., Sander et al., Verh\"ulsdonk et al., and Kong et al.~\cite{khan, sander, verhulsdonk2024shape, kong} explored DeepSDF-based representations for cardiac and vascular structures. In particular, Kong et al.~\cite{kong} introduced positional encodings and representations of congenital heart defects. Building on these ideas, Yuan and Sørensen~\cite{yuan20234d, sorensen2024spatio} proposed spatiotemporal SDF reconstruction for the heart and aorta. More recently, Qiao et al.~\cite{qiao2025personalized} proposed a transformer-based mesh encoder-decoder architecture for modeling spatiotemporal cardiac data.

Segmentation and surface reconstruction methods have also been extended with uncertainty quantification (UQ). Chang et al.~\cite{chang2011efficient} proposed efficient MCMC sampling with implicit shape representations, while more recent work has focused on stochastic surface reconstruction, including stochastic Poisson methods~\cite{sellan2022stochastic}, neural stochastic screened Poisson~\cite{sellan2023neural}, and earlier work quantifying uncertainty in point-cloud surfaces~\cite{pauly2004uncertainty}. For a recent overview of point-cloud-based medical shape modeling, we refer to~\cite{zhang2025survey}.

Statistical shape modeling (SSM) is widely used in cardiac shape reconstruction and can also be employed for UQ. The DeepSSM framework~\cite{bhalodia2018deepssm} introduced latent-space learning of population-level shape variability. Subsequent extensions include Uncertain-DeepSSM~\cite{adams2020uncertain}, vib-DeepSSM~\cite{adams2022images}, and the Bayesian BVIB-DeepSSM~\cite{adams2023fully}, which explicitly encode uncertainty in learned representations.

Sampling-based methods such as Markov chain Monte Carlo (MCMC) have proven highly effective for uncertainty propagation in medical modeling and inference. In particular, Hamiltonian Monte Carlo (HMC)~\cite{neal2011mcmc} and its adaptive variant, the No-U-Turn Sampler (NUTS)~\cite{hoffman2014no}, have demonstrated superior efficiency in exploring high-dimensional posterior distributions. In cardiovascular modeling, Du and Wang~\cite{du2022reducing} developed a deep learning-assisted parallel-chain MCMC method to reduce geometric uncertainty in patient-specific hemodynamics, highlighting the utility of Bayesian inference in cardiac reconstruction. More broadly, Bayesian MCMC approaches have been emphasized as essential for reproducibility in scientific modeling~\cite{volodina2021importance}. Alternatives such as deep ensembles have also been proposed to approximate posterior uncertainty without expensive sampling~\cite{dangelo}.

\section{Preliminaries}
In \cref{sec:deepsdf}, we will describe the Deep Signed Distance (DeepSDF) framework for cardiac shape reconstruction as in~\cite{verhulsdonk2024shape} and in \cref{sec:mcmc}, we will recap the idea of Markov Chain Monte Carlo methods.

\subsection{Shape Atlas generation via Deep Signed Distance Function (DeepSDF)}
\label{sec:deepsdf}
In the context of Deep Signed Distance Functions we describe shapes implicitly as the zero level $f_\theta^{-1}(0)$ set of a learned signed distance function $f_\theta$ with parameters $\theta$.
We train a latent code representation $\latentcode_i$ for every patient $i$ that encodes the joint representation of multiple cardiac shapes $S_{\ell,i}$ for $\ell=1,\ldots,L$. In the context of this paper, we restrict ourselves to $L=4$ and learn representation for the left- and right ventricle endo- end epicardia.
For the training process, we assume samples $X_i \coloneqq \left(\vx_k, \sdfvec_k \right)_{k=1}^{K_i}$ used for learning are $K_i$ pairs of spatial coordinates $\vx_k$ with the signed distance vector $\sdfvec_k\in\R^L$ for the $i$-th biventricular shape.
The set of all $N$ anatomical bi-ventricular samples is denoted as $X \coloneqq \left( X_i \right)_{i=1}^N$.
We associate a coupled latent code representation $\latentcode_i \in \latentspace$ for every anatomical shape $i$ and denote the set of latent codes as $Z\coloneqq \left( \latentcode_i \right)_{i=1}^N$.
In training, we minimize the mismatch between the sampled signed distances and the ones estimated by the network $\shapeopmulti$ and assume a zero mean Gaussian prior distribution with covariance $\sigma^2I$ on the latent codes, which gives rise to the loss term
\begin{equation}\label{eq:train_loss}
    \mathcal{L}(\nnweights, Z) = \frac{1}{N} \sum_{\substack{i=1,\ldots,N \\(\vx_k, \sdfvec_k) \in X_i}} \frac{1}{4 K_i} \lVert \shapeopmulti(\vx_k, \latentcode_i) - \sdfvec_k\rVert^2 + \frac{1}{\sigma^2} \Vert\latentcode_i\Vert^2_2.
\end{equation}
For each layer an additional weight $c_i$ is introduced such that $\text{softplus}(c_i) = \ln(1+e^{c_i})$ serves as an upper bound $\text{softplus}(c_i) \ge \Vert W_i \Vert_p$ for the Lipschitz constant.
Integrating this Lipschitz-regularization into our previous loss-functional $\mathcal{L}$ leads to our finally used cost-functional 
\begin{equation}
J(\nnweights, Z) = \mathcal{L}(\nnweights, Z) + \alpha \prod_{c_i \in C(\theta)} \text{softplus}(c_i),\label{eq:lipschitz_nn_loss}
\end{equation}
where $C(\theta) = (c_i)_{i=1}^M$ denotes the network parameter dependent per-layer Lipschitz bounds.

\subsection{Shape Inference from Point Clouds}
After learning the SDF network, a new anatomical bi-ventricular shape can be inferred from sparse point clouds of any combination of surfaces or given signed distances.
For this, we consider $K$ given samples consisting of triplets~$\tilde{Y}$ of spatial coordinates $\vx_k \in \R^3$, a single signed distance $s_k \in \R$, and an index of the surface $j_k \in \{1, 2, 3, 4\}$.
If a point $\vx_k$ lies on the surface $j_k$, then $s_k = 0$.
Finding the bi-ventricular reconstruction of a point cloud thus reduces to finding its latent code representation $\latentcode$ by minimizing the following problem
\begin{equation}
    \min_{\latentcode} \frac{1}{K} \sum_{(\vx_k, s_k, j_k) \in \tilde{Y}} \left( \left( \shapeopmulti (\vx_k, \latentcode ) \right)_{j_k} - s_k \right)^2 + \frac{1}{\sigma^2} \Vert \latentcode \Vert_2^2.
    \label{eq:inference_loss}
\end{equation}

\subsection{Markov Chain Monte Carlo Methods}
\label{sec:mcmc}
For uncertainty quantification in various applications, computing the posterior distribution is often a key objective. However, directly computing the posterior is computationally prohibitive in many scenarios, especially when dealing with high-dimensional models. To address this challenge, sampling methods, such as Metropolis-Hastings-type Markov Chain Monte Carlo (MCMC) algorithms, are frequently employed.

For Bayesian inference we aim to sample from the posterior
\begin{equation*}
\pi(z) \propto \exp(-\Phi(z)), \quad \text{with } 
\Phi(z) = -\log p(b \mid z) - \log p(z),
\end{equation*}
where $\Phi$ consists of the negative log-likelihood $-\log p(b\mid z)$ and the negative logarithm of the prior $-\log p(z)$. In our case the observation $b$ is given by the considered point cloud, whereas $z$ can be given by the latent variable $\mathbf{z}$ or $(\mathbf{z},\zeta)$ if the noise level $\zeta$ is also sampled (c.f. \Cref{sec:noise_inference}).
In settings where the direct computation of the posterior is infeasible, one can often use Markov Chain Monte Carlo (MCMC) methods to construct a Markov chain with stationary distribution~$\pi$. A generic Metropolis--Hastings update proposes a new point $z' \sim q(\cdot \mid z_k)$ and accepts it with probability
\begin{equation*}
\alpha(z_k,z')=\min\!\left(1,\;
\frac{\pi(z')\, q(z_k\mid z')}{\pi(z_k)\, q(z'\mid z_k)}\right).
\end{equation*} 
Hamiltonian Monte Carlo~\cite{neal2011mcmc} (HMC) introduces a momentum variable $p \sim \mathcal N(0,M)$ and defines the Hamiltonian
\begin{equation*}
H(z,p) = \Phi(z) + \tfrac12 p^\top M^{-1}p .
\end{equation*}
A proposal is obtained by simulating Hamilton's equations
\begin{equation*}
\dot z = M^{-1}p, \qquad
\dot p = -\nabla \Phi(z)\,,
\end{equation*}
using a symplectic integrator (e.g.\ leapfrog) with step size $\epsilon$ for $L$ steps.  
The proposal $(z',p')$ is accepted with probability
\[
\alpha = \min\left(1,\; e^{-H(z',p') + H(z_k,p_k)}\right).
\]

The No-U-Turn Sampler~\cite{hoffman2014no} (NUTS) adapts the trajectory length automatically. It recursively builds a binary tree of leapfrog steps in forward and backward time and stops expansion when the
trajectory begins to reverse direction, detected by the ``no U-turn'' criterion
\[
(z^+ - z^- )^\top M^{-1} p^- < 0
\quad \text{or} \quad
(z^+ - z^- )^\top M^{-1} p^+ < 0,
\]
ensuring efficient exploration without manual tuning of~$L$.

\section{Bayesian Inference for DeepSDF}
In this section, we look at the shape inference task from a stochastic point of view, resulting in a Bayesian reformulation that allows for uncertainty quantification. 
Let us assume we have a ground truth shape described via a signed distance function $S_{gt}(x) = 0$ and that the measurement points are generated by applying Gaussian noise with variance $\zeta^2$ to the surface point $\hat x_k$ in the normal direction $n(x_k)$ with $\|n(x)\| = 1$ for all surface points~$x$. We obtain measurement points 
\begin{equation}\label{eq:noise}
    y_k \sim x_k + n(x_k)\,\mathcal{N}\!\big(0,\,\zeta^2\big)\,.
\end{equation}
For small curvature, the resulting signed distance value at the point $y_k = x_k + \tau_k n(x_k)$ can be computed via the Taylor approximation
\begin{equation}
    S_{gt}(y_k) \approx S_{gt}(x_k) + \tau_k\, n(x_k)^\top \nabla S_{gt}(x_k)\,.
\end{equation}
For a signed distance function we have $\|\nabla S_{gt}(x)\| = 1$ for all $x$, and on the surface the gradient equals the normal direction, i.e.\ $\nabla S_{gt}(x_k) = n(x_k)$. We thus obtain
\begin{equation}
    S_{gt}(y_k) \approx S_{gt}(x_k) + \tau_k \|n(x_k)\|^2 = S_{gt}(x_k) + \tau_k\,.
\end{equation}
For the sampled points, we can therefore use the approximation 
\begin{equation}
    s_k \sim \mathcal{N}\!\big(0,\,\zeta^2\big)\,,
\end{equation}
i.e.\ the signed distance is distributed according to a one-dimensional Gaussian. 
For the case of points sampled from the true surface, this is equivalent to stating that the model prediction $\shapeopmulti(\vx_k, \latentcode)_{j_k}$ is sampled from the same Gaussian. 

For the general case, the likelihood of a point cloud $\tilde{Y} = \{(\vx_k, s_k, j_k)\}_{k=1}^K$ given the latent code $\latentcode$ is
\begin{equation}
    p(\tilde{Y} \mid \latentcode) \;\propto\;
    \exp\!\left(
        - \frac{1}{2K\zeta^2}
        \sum_{(\vx_k, s_k, j_k) \in \tilde{Y}}
        \big( (\shapeopmulti(\vx_k, \latentcode))_{j_k} - s_k \big)^2
    \right).
\end{equation}
Taking the negative log-likelihood yields the quadratic error term in~\cref{eq:inference_loss}. 

To regularize the solution, we place a Gaussian prior on the latent codes
\begin{equation}
    \latentcode \sim \mathcal{N}(\mu, \tilde{\sigma}^2 \mathbf{I})\,.
\end{equation}
Combining likelihood and prior via Bayes’ theorem gives the posterior distribution
\begin{equation}
    p(\latentcode \mid \tilde{Y}, \theta) \;\propto\; 
    p(\tilde{Y} \mid \latentcode, \theta)\, p(\latentcode)\,.
\end{equation}
The negative log-posterior is given, up to additive constants, as
\begin{align}\label{eq:map}
    -&2 \log p(\latentcode \mid \tilde{Y}, \theta) \\
    &= \frac{1}{K\zeta^2}
    \sum_{(\vx_k, s_k, j_k) \in \tilde{Y}}
        \left( (\shapeopmulti(\vx_k, \latentcode))_{j_k} - s_k \right)^2
        + \frac{1}{\tilde{\sigma}^2} \|\latentcode - \mu\|_2^2.\nonumber
\end{align}
For $\zeta^2 = 1$, $\mu = 0$, and $\tilde\sigma = \sigma$, we obtain the same functional that is used for the inference problem in \cref{eq:inference_loss}. In other words, the loss function in \cref{eq:inference_loss} can be interpreted as the \emph{negative log-posterior}. Minimizing it yields the maximum a posteriori (MAP) estimate of the latent code, while sampling from it via MCMC allows us to approximate the full posterior distribution.

With this model, we leverage MCMC methods to sample latent codes of the posterior that encode continuous SDF representations of multiple cardiac surfaces. We can statistically evaluate the generated samples on arbitrary meshes. Natural choices include the nodes of the MAP reconstruction or a uniform grid that covers the entire domain. In particular, we can evaluate the information at arbitrarily fine resolution.

\subsection{Implementation and training}
In this section, we describe the implementation and training of the DeepSDF forward model as well as the probabilistic statistical model. Following the approach of Verhulsdonk et al.~\cite{verhulsdonk2024shape}, we employ a fully connected neural network comprising five layers with 256 neurons each and \texttt{tanh} activation functions. The network maps the concatenation of a spatial coordinate and a 64-dimensional latent code to an $L$-dimensional vector of signed distance function (SDF) values, where $L=4$, corresponds to the endocardial and epicardial surfaces of the left and right ventricles. To enhance reconstruction stability, we apply Lipschitz regularization following Liu et al.~\cite{liu_learning_2022}.

The model is trained using 40 meshes from the dataset of Grandits et al.~\cite{grandits_public_2024}, which builds upon the publicly available cardiac geometries of Strocchi et al.~\cite{strocchi_publicly_zenodo_2020} and Rodero et al.~\cite{rodero_virtual_2021}. The remaining four meshes are reserved for evaluation and further tests of our method. 

We train the model on the loss~\eqref{eq:train_loss} with the Adam optimizer~\cite{kingma_2014_adam} for 2000 epochs with a learning rate of 0.005 and two learning rate reductions with the factor $0.2$ after 1800 and 1950 epochs. As in~\cite{verhulsdonk2024shape} we set $1/\sigma^2=1.8\times10^{-8}$ and $\alpha=1.9\times 10^{-6}$. For the mesh generation of a sample (or MAP estimate) we use a $128^3$ grid for signed distance function evaluation and compute the zero level set. 

The statistical model is implemented using the Pyro probabilistic programming library~\cite{bingham2019pyro}, which is built on PyTorch and provides a flexible framework for stochastic variational inference and MCMC-based sampling.

\subsection{Quantitative Evaluation}

To assess the quality of the generated reconstructions and their associated uncertainty estimates, we adopt an evaluation strategy inspired by Vasconcelos et al.~\cite{vasconcelos}, who proposed similar metrics for image-based implicit representations. In our setting, reconstructions are evaluated on meshes generated on the zero level set of an MAP estimate. On every node of the mesh, we compare the distribution of signed distance values among the samples of the chain against the true signed distance to a ground-truth mesh and statistically evaluate the coverage. 

Given a set of latent codes $\{\mathbf{z}_i\}_{i=1}^N$ and mesh coordinates $\{\mathbf{x}_k\}_{k=1}^K$, we compute the empirical distribution of signed distance function (SDF) values at each coordinate:
\[
    F_N(\mathbf{x}_k) = \{\mathbf{S}_\theta(\mathbf{x}_k,\mathbf{z}_i)\}_{i=1}^N.
\]
For reference, we define $f^\ast(\mathbf{x}_k)$ as the implicit distance from $\mathbf{x}_k$ to the GT mesh, representing the true SDF value at that coordinate.  

For a quantile level $0 \leq q \leq 1$, we compute the $q$-quantile interval for every mesh coordinate distribution $(q^k_l,q^k_r)=(Q^k_{0.5-q/2},Q^k_{0.5+q/2})$ where $Q^k_t$ denotes the $t$ quantile of the empirical distribution $F_N(x_k)$. For given $q$ and $k$ we can evaluate if the true sdf value lies within the $q$-quantile of $F_N(x_k)$. With
\begin{align*}
    C_{F_N,f^\ast,q}(x_k) = \begin{cases} 1 &\text{if } q^k_l \leq f^\ast(x_k)\leq  q^k_r \\
    0 &\text{else}
    \end{cases}\,,
\end{align*}
we can define the \emph{achieved coverage} (AC) as
\begin{equation}\label{eq:ac}
    \operatorname{AC}(f^\ast,F_N,q) = \frac{1}{K}\sum_{k=1}^K C_{F_N,f^\ast,q}(x_k)\,.
\end{equation}
Ideally, the achieved coverage matches the nominal coverage $q$, i.e.\ $\operatorname{AC}(f^\ast,F_N,q) \approx q$. Note that this quantity is sometimes referred to as Prediction Interval Coverage Probability (PICP)~\cite{bazionis2021review}

To quantify deviations from perfect calibration, we compute the \emph{expected calibration error} (ECE) across a set of $M$ evenly spaced quantile levels $\{q_m\}_{m=1}^M$ as
\begin{equation}\label{eq:ece}
    \operatorname{ECE}(f^\ast,F_N) = \frac{1}{M} \sum_{m=1}^M \big|\operatorname{AC}(f^\ast,F_N,q_m) - q_m\big|.
\end{equation}

\section{Numerical Results}
In this section, we describe numerical experiments for the uncertainty quantification of the above MCMC DeepSDF framework.

\subsection{Partial Data Availability}

\begin{figure}
    \input{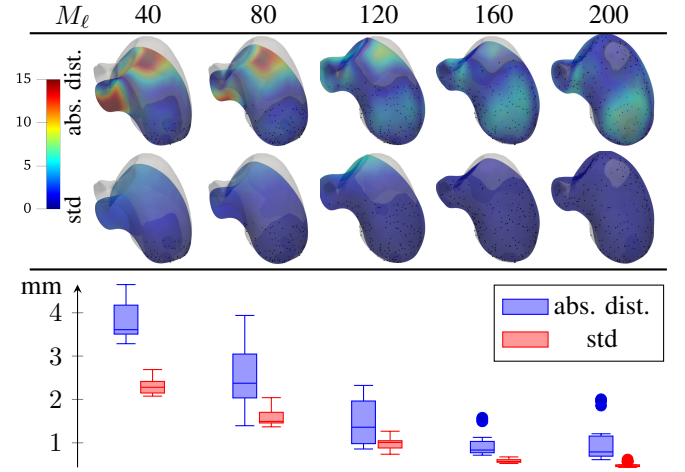}
    \caption{Effect of increasing the number of input points on reconstruction quality and uncertainty estimates.}
    \label{fig:vert_series}
\end{figure}

As a first experiment, we test the proposed method in a scenario with incomplete data, i.e., we evaluate its performance on point clouds with varying degrees of sparsity. \Cref{fig:vert_series} illustrates the effect of increasing the number of available input points in the vertical direction. In detail, we consider a dense sample of points $(\mathbf{x}_i)_{i=1}^M$ from the surface of the left ventricle of hearts from the test set. For $\ell=1,\ldots,5$, we then sample an increasing number of points $M_\ell=20\ell$ with $y$-coordinate in the expanding range $(y_{\operatorname{min}},(\ell /5)y_{\operatorname{max}})$, where $y_{\operatorname{min}}$ and $y_{\operatorname{max}}$ denote the minimum and maximum values of the $y$-coordinates of the initial point cloud. We generate a total of $N=4000$ samples $(z_i)_{i=1}^N$ using the No-U-Turn Sampler with 20 parallel chains and evaluate the resulting reconstruction errors and uncertainty quantification on the reconstruction based on the MMSE estimator $\hat{z}=\frac{1}{N}\sum_{i=1}^Nz_i$. We repeat this experiment five times for all four test meshes with differently sampled point clouds and compute for every resulting MMSE reconstruction the pointwise average of the absolute distance to the ground-truth mesh and the variance of the sdf values. We show boxplots of these quantities (Fig.~\ref{fig:vert_series}) for one run on mesh 23.
As anticipated, reconstructions become more accurate and exhibit reduced uncertainty as the number of points increases. The top panel depicts the absolute distance between the reconstructed and reference shapes, whereas the bottom panel shows the standard deviation, reflecting the uncertainty in the reconstructions across different point cloud densities.

\subsection{Comparison of Laplace approximation and MCMC samplers}
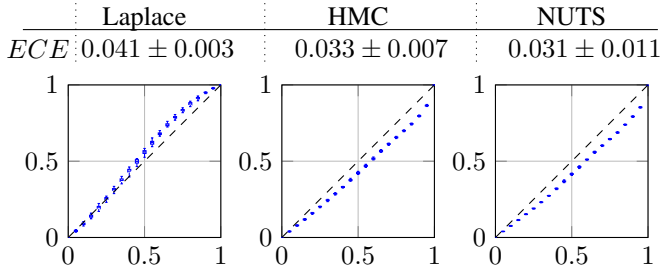
\begin{figure}
    \centering
\begin{tikzpicture}
\node at (-0.4,2.5) {$ECE$};
\draw[black] (-0.6,2.75) -- (7.8,2.75);

\draw[dotted] (0.1,3.1) -- (0.1,2.3);
\draw[dotted] (2.6,3.1) -- (2.6,2.3);
\draw[dotted] (5.4,3.1) -- (5.4,2.3);

  \begin{axis}[
    width=3.6cm, height=3.6cm,
    xmin=0, xmax=1, ymin=0, ymax=1,
    title={ \Longstack{Laplace\\\quad$0.041\pm0.003$}},
    xtick={0,0.5,1},
    ytick={0,0.5,1},
    grid=both,
    boxplot/draw direction=y,
    every boxplot={
      box extend=0.02        
    },
    scaled x ticks=false,
    scaled y ticks=false
  ]

  \pgfplotstableread[col sep=comma]{data/results_Hessian_short.csv}\rawtable
  \pgfplotstabletranspose\transtable{\rawtable}

  \pgfplotstablegetcolsof{\transtable}
  \pgfmathtruncatemacro{\ncols}{\pgfplotsretval-1} 

  \foreach \col in {0,...,21} {
    \pgfmathsetmacro{\xcoord}{\col/20-0.05}
    \addplot+[
      boxplot,
      boxplot/draw position=\xcoord,
      draw=blue,
      boxplot/box extend=0.02, 
      mark={}, 
      fill opacity=0.3
    ] table[y index=\col] {\transtable};
  }

  \addplot[domain=0:1, samples=2, black, dashed] {x};

  \end{axis}

  \begin{axis}[
    at={(140,0)},
    width=3.6cm, height=3.6cm,
    xmin=0, xmax=1, ymin=0, ymax=1,
    title={\Longstack{HMC\\\quad$0.033\pm0.007$}},
    xtick={0,0.5,1},
    ytick={0,0.5,1},
    grid=both,
    boxplot/draw direction=y,
    every boxplot={
      box extend=0.02        
    },
    scaled x ticks=false,
    scaled y ticks=false
  ]

  \pgfplotstableread[col sep=comma]{data/results_nuts_zeta_2_short.csv}\rawtable
  \pgfplotstabletranspose\transtable{\rawtable}

  \pgfplotstablegetcolsof{\transtable}
  \pgfmathtruncatemacro{\ncols}{\pgfplotsretval-1} 

  \foreach \col in {0,...,21} {
    \pgfmathsetmacro{\xcoord}{\col/20-0.05}
    \addplot+[
      boxplot,
      boxplot/draw position=\xcoord,
      draw=blue,
      boxplot/box extend=0.02, 
      mark={}, 
      fill opacity=0.3
    ] table[y index=\col] {\transtable};
  }

  \addplot[domain=0:1, samples=2, black, dashed] {x};

  \end{axis}

  \begin{axis}[
    at={(280,0)},
    width=3.6cm, height=3.6cm,
    xmin=0, xmax=1, ymin=0, ymax=1,
    title={\Longstack{NUTS\\\quad$0.031\pm0.011$}},
    xtick={0,0.5,1},
    ytick={0,0.5,1},
    grid=both,
    boxplot/draw direction=y,
    every boxplot={
      box extend=0.02        
    },
    scaled x ticks=false,
    scaled y ticks=false
  ]

  \pgfplotstableread[col sep=comma]{data/results_nuts_zeta_3_short.csv}\rawtable
  \pgfplotstabletranspose\transtable{\rawtable}

  \pgfplotstablegetcolsof{\transtable}
  \pgfmathtruncatemacro{\ncols}{\pgfplotsretval-1} 

  \foreach \col in {0,...,21} {
    \pgfmathsetmacro{\xcoord}{\col/20-0.05}
    \addplot+[
      boxplot,
      boxplot/draw position=\xcoord,
      draw=blue,
      boxplot/box extend=0.02, 
      mark={}, 
      fill opacity=0.3
    ] table[y index=\col] {\transtable};
  }

  \addplot[domain=0:1, samples=2, black, dashed] {x};

  \end{axis}

\end{tikzpicture}
    \caption{Quantitative comparison of different sampling methods. We ran each method 20 times and show boxplots of the computed $AC$ for 20 quantiles, as well as the mean and standard deviation of the $ECE$.}
    \label{fig:comparison_sampling}
\end{figure}

We compare the uncertainty quantification results of our proposed approach generated by different sampling strategies, including Laplace's approximation of the posterior (Laplace) and sampling with the Hamiltonian Monte Carlo sampler (HMC) and the No-U-Turn Sampler (NUTS). 
For Laplace approximation, the posterior distribution from~\eqref{eq:map} is approximated by a Gaussian distribution that can be sampled directly.

Because MAP optimization may converge to different local minima depending on initialization, we sample from distributions centered at multiple local optima to capture variability across the latent space. For comparison with MCMC-based methods, we perform quantitative evaluations on 20 distinct MAP estimates and analyze the distribution of the $\operatorname{AC}$ from~\eqref{eq:ac} for quantiles spaced by $0.05$ as well as the mean expected calibration error (ECE) from~\eqref{eq:ece}.

For both HMC and NUTS, we run 20 chains in parallel using $100$ warm-up steps and $500$ samples each. For both methods, we take an initial step size of $1$ and we adapt the step size and mass matrix during warm-up using Dual Averaging and the Welford scheme, respectively. We use a target acceptance probability of $0.8$. For NUTS, we apply multinomial sampling as described in~\cite{betancourt2017conceptual}. For this comparison we chose $\zeta^2=1$ in the likelihood.

\Cref{fig:comparison_sampling} presents the results obtained using Hessian-based sampling in comparison with MCMC approaches, specifically Hamiltonian Monte Carlo (HMC) and the No-U-Turn Sampler (NUTS).
All sampling strategies provide well-calibrated uncertainty quantification when evaluated on the MAP estimates. Laplace's approximation is computationally the most efficient, since it relies solely on the computation of the MAP and the Hessian of the posterior~\eqref{eq:map}. The MAP estimate can be computed using the Adam optimization algorithm on GPU hardware within a few minutes and the Hessian can be computed efficiently via automatic differentiation. The iterative sampling strategies are computationally more demanding although multiple chains can be run in parallel. 

Regarding computational cost, we provide a comparison for a single chain with different sample sizes in~\Cref{tab:runtime}. For the MCMC approaches, we use 50 warm-up steps and report the average effective sample size (ESS), the resulting runtime, and the number of function evaluations (all including the warm-up phase). For the Laplace approximation, we report the runtime and the number of function evaluations, which arise primarily from the MAP estimation procedure. The effective sample size (ESS) is estimated based on the sum of autocorrelations, following the standard initial positive sequence estimator used in modern MCMC diagnostics.

\begin{table}[t]
\centering
\caption{\label{tab:runtime}Comparison of samplers for different sample sizes.}
\begin{tabular}{l c c c c}
\toprule
Sampler & \#Samples & Mean ESS & Runtime [s] & Fun Evaluation \\
\midrule
Laplace  &  &  & 164 & 20000 \\
\midrule
HMC  & 50  & 59.87 & 48  & 8149 \\
HMC  & 100 & 252.62 & 62 & 14249 \\
HMC  & 500 & 312.51 & 347 & 63049 \\
\midrule
NUTS & 50  & 36.16 & 47 & 5625 \\
NUTS & 100 & 62.86 & 57 & 7591 \\
NUTS & 500 & 303.12 & 191 & 24055 \\
\bottomrule
\end{tabular}
\end{table}

\subsection{Multichamber Uncertainty}

\begin{figure*}
    \centering
    \input{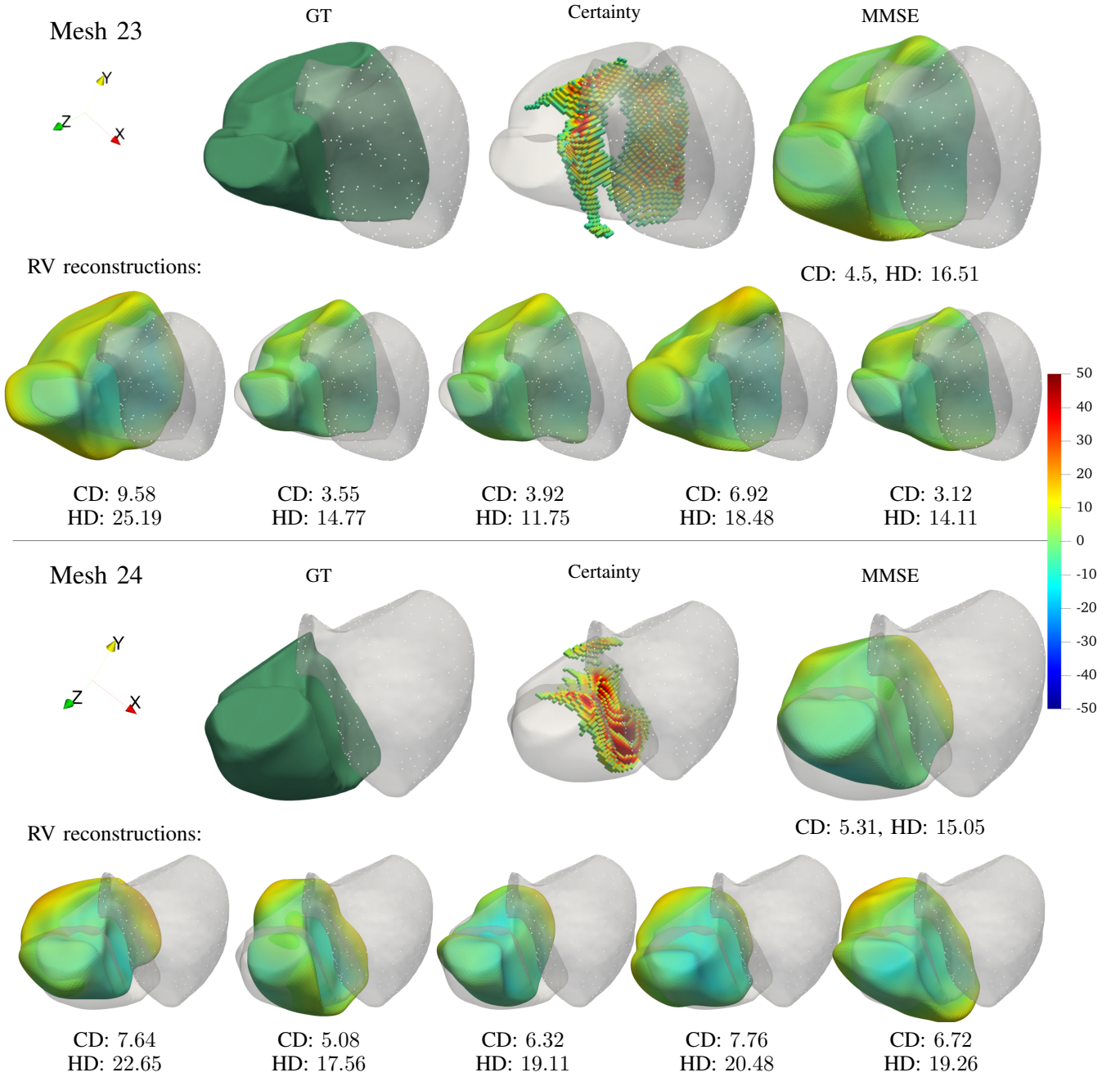}

    \caption{The Maximum Likelihood Estimator based on LV points can produce multiple variable RV shapes. A certainty quantification helps to get reliable cross-chamber information. For the RV reconstructions and the SSME reconstruction we colorcode the implicit distance and provide CD and HD w.r.t. to the ground truth mesh.}
    \label{fig:multichamber}
\end{figure*}

The proposed UQ-framework can also be leveraged to gain cross-surface information in a multimodal setting such as ours. Since we use a joint prior for multiple surfaces we can evaluate uncertainty information on all signed distance functions based on samples generated from point clouds of a different surface.
As illustrated in \Cref{fig:multichamber}, the maximum a posteriori estimator (MAP) based solely on LV point data can produce multiple variable reconstructions of the right ventricle (RV), depending on the initialization of the optimization process. By incorporating uncertainty estimation, our framework can infer regions that are most likely to contain sdf values that are close to zero and therefore are more likely to belong to the surface. This provides more robust cross-chamber information despite the ambiguity in RV reconstructions.

The uncertainty visualization in~\Cref{fig:multichamber} is obtained by evaluating MCMC-generated samples on a fine uniform 3D grid based on 200 LV points for two meshes from the test set. For each grid point, we evaluate the RV signed distance function and count occurrences where its absolute value is smaller than the mesh resolution. The resulting certainty map represents a threshold of a voxelized count of these occurrences, showing regions belonging to the RV with high certainty. In both test cases, the voxels with high certainty actually correspond to the surface part adjacent to the left ventricle, which seems reasonable. We provide the MMSE reconstruction and multiple MAP reconstructions of the right ventricle based on the left ventricle point cloud, together with their Chamfer distance (CD) and Hausdorff distance (HD) to the ground truth, to illustrate the variability in this setting.

\subsection{Comparison with Other Shape UQ Methods}

\begin{figure*}[ht]
    \centering
    \input{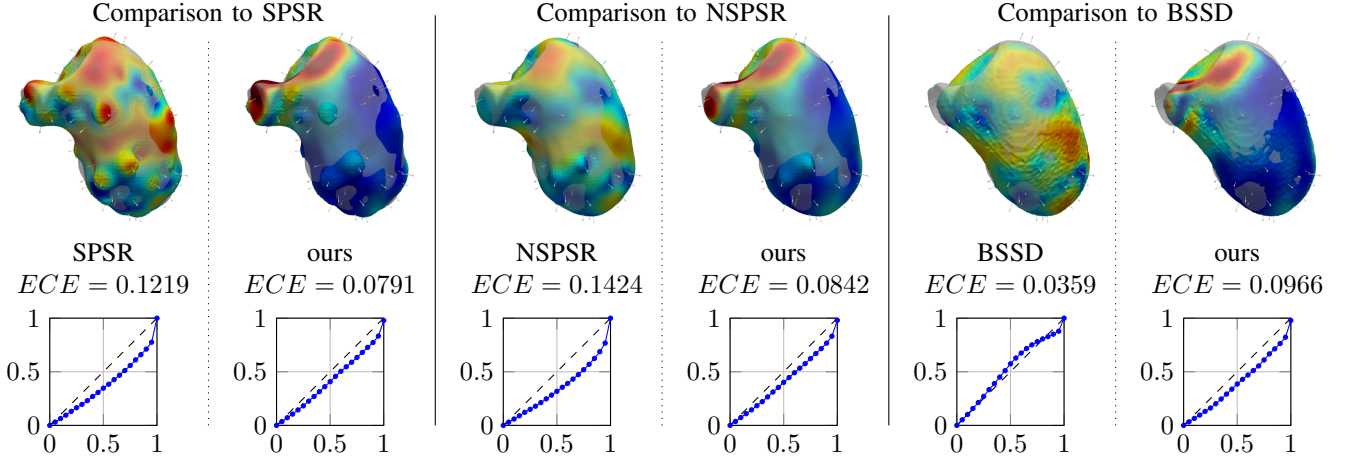}
    \caption{Comparison with SPSR (left), NSPSR (middle), and BSSD (right). We evaluated our samples on the reconstructions produced by the other methods and show the computed $AC$ and $ECE$ for 20 quantiles.}
    \label{fig:comparison}
\end{figure*}

We further compare our approach with three recent methods for uncertainty-aware surface reconstruction: stochastic Poisson surface reconstruction (SPSR)~\cite{sellan2022stochastic}, neural stochastic Poisson surface reconstruction (NSPSR)~\cite{sellan2023neural}, and Bayesian Smooth Signed Distance (BSSD)~\cite{pujol2025bayesian}. All methods provide uncertainty estimates for the reconstructed surfaces, but rely on different mechanisms. 

A key advantage of our framework is that it does not require surface normal information in the input point cloud which are required by the compared methods. This independence from normal vectors enhances the applicability of our method in real-world clinical settings, where such information may not be available. A direct quantitative comparison with these baselines is challenging because their uncertainty estimates inherently depend on the quality of the initial surface reconstruction, which differs across methods. BSSD yields smoother reconstructions, which may affect pointwise uncertainty calibration.

Since our method can perform uncertainty quantification on arbitrary meshes, we can compare the uncertainty quantification on the reconstructions of the other methods. However, evaluating uncertainty at reconstruction points based on the sampled signed distances is problematic at points farther from the zero level sets. All other methods provide pointwise variance estimates, which allow us to compute quantiles for comparison with the true SDF values at each node using the $AC$ and $ECE$ metrics defined in~\eqref{eq:ac} and~\eqref{eq:ece}, respectively. \Cref{fig:comparison} illustrates the comparison of reconstructions obtained using SPSR (left), NSPSR (middle) and BSSD (right). The SPSR approach appears less effective in capturing reconstruction uncertainty according to the $AC$ and $ECE$ metrics; likely, a denser point cloud would be required. Both NSPSR and our method successfully identify regions of greater uncertainty.

\subsection{Inferring the Noise Parameter}\label{sec:noise_inference}

\begin{figure*}
    \centering
    \input{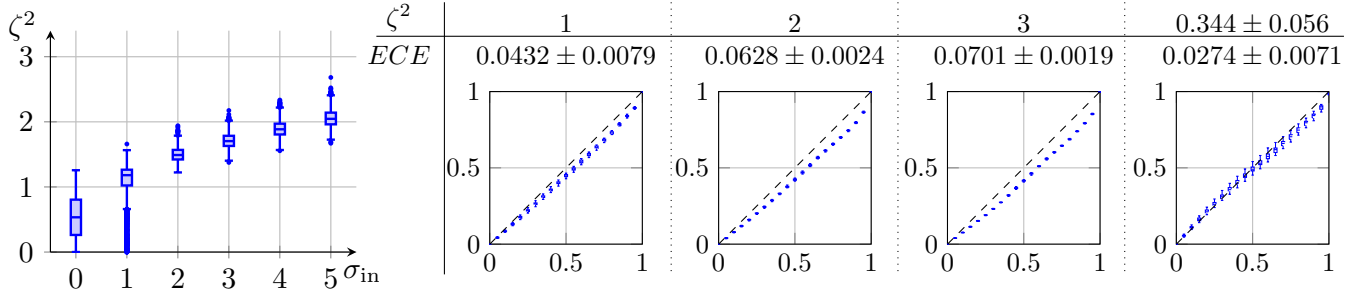}
    \caption{Left: inferred $\zeta^2$ for different levels of input noise $\sigma_{\text{in}}$. Right: quantitative comparison across different values of $\zeta^2$. We run each method 20 times and present boxplots of the computed $AC$ over 20 quantiles, along with the mean and standard deviation of the $ECE$. In the final experiment, $\zeta^2$ is jointly sampled with the latent variables.}
    \label{fig:inf_noise}
\end{figure*}

A further application of our framework is joint inference of the observation noise parameter when the noise level is unknown. In our probabilistic model, we assume Gaussian observation noise $y_k \sim x_k + n(x_k)\,\mathcal{N}\!\big(0,\,\zeta^2\big)$, where $\zeta^2$ denotes the noise variance (cf.~\eqref{eq:noise}). This parameter can either be fixed when the noise level is known or treated as an additional variable inferred during the MCMC sampling process, thereby enabling the model to adapt to varying noise conditions in the data.

\Cref{fig:inf_noise} illustrates the inferred noise parameter $\zeta^2$, sampled from the uniform prior $\zeta^2 \sim \mathcal{U}(0,10)$, for different levels of injected input noise on 100 LV points. Although the exact noise variance cannot be recovered perfectly, the inferred values of $\zeta^2$ exhibit a strong correlation with the true noise levels, demonstrating the model’s ability to capture relative noise magnitudes. We additionally evaluate the effect of an inferred $\zeta^2$ on the $ECE$ metric and obtain better calibrated results for this procedure (cf.~\Cref{fig:inf_noise}).

\section{Discussion}

We showed that reformulating the original inference problem within a Bayesian framework enables the use of Markov Chain Monte Carlo (MCMC) sampling techniques to obtain well-calibrated uncertainty quantification for cardiac shape reconstruction. The resulting posterior samples can be evaluated in a mesh-independent manner and provide uncertainty information for all cardiac chambers represented by the prior. In particular, when data availability is sparse, prior-based sampling strategies yield more reliable cross-chamber information than standard maximum a posteriori (MAP) estimation, which can result in various reconstruction based on the same input depending on the optimization.

A multimodal Laplace approximation of the prior offers a computationally efficient way to access uncertainty information after shape reconstruction, enabling rapid approximations of posterior variability. In contrast, Hamiltonian MCMC methods deliver more accurately calibrated posterior samples at the cost of substantially higher computational effort. For real-time clinical applications, iterative sampling approaches remain largely prohibitive due to their runtime requirements, whereas the Laplace approximation can provide reliable uncertainty estimates within minutes, making it a more practical option in time-critical settings.

\subsection{Limitations and future directions.}
There are some limitations related to our approach:

First, balancing the loss terms in \eqref{eq:map} requires careful selection of the measurement noise variance $\zeta$ and the parameters of the prior distribution $\mathcal{N}(\mu,\Sigma)$, introducing additional design choices. In this work, we assume $\Sigma = \tilde{\sigma}^2 \mathbf{I}$ and select the parameters such that the original loss function in \eqref{eq:inference_loss} is recovered. Alternatively, $\mu$ and $\Sigma$ could be estimated from the latent codes of the training shapes using maximum likelihood estimators, enabling a more data-driven prior.

Second, the expressiveness of the shape prior is limited by the size and diversity of the training data. Training on larger-scale datasets could yield more expressive shape priors, potentially improving reconstruction accuracy and uncertainty estimation, particularly in complex anatomical regions. Moreover, uncertainty quantification based on a DeepSDF shape prior could be extended to other anatomical structures and cardiac reconstruction tasks, such as modeling the aorta.

Finally, although inference of the noise level from point cloud data does not yield exact estimates, our experiments indicate that the proposed sampling strategies can recover meaningful information about the underlying noise level. This suggests that Bayesian sampling approaches may offer a promising direction for improved noise characterization in future work.

\bibliographystyle{ieeetr}
\bibliography{references}

\end{document}